\documentclass[a4paper,11pt]{article}
\usepackage{listings}
  \usepackage{mathrsfs}
       \usepackage[mathscr]{euscript}
 \newcommand{\red}[1]{\textcolor{red}{#1}}

\usepackage{vmargin}
\setmarginsrb{1in}{1in}{1in}{1in}{0pt}{0pt}{0pt}{6mm}
\usepackage{amsthm,amsmath,amssymb}

\usepackage{mathtools}

\usepackage{xcolor}
\date{}

\definecolor{bblue}{rgb}{0.1,0.4,0.6}

\newcommand{\OO}{\mathcal{O}}

\newcommand{\np} {{\sf NP}}
	\newcommand{\fpt} {{\sf FPT}\xspace}
	\newcommand{\nph} {{\sf NP}-hard\xspace}
	\newcommand{\npc} {{\sf NP}-complete\xspace}
	\newcommand{\woh} {{\sf W}$[1]$-hard\xspace}
	\newcommand{\woc} {{\sf W}$[1]$-complete\xspace}

\usepackage{tcolorbox}
\tcbuselibrary{skins}
\usepackage{xcolor}
\colorlet{mix}{red!50!black}

\usepackage{amsmath}
\usepackage{amsthm}

\usepackage{nameref}
\definecolor{ForestGreen}{rgb}{0.1333,0.5451,0.1333}
\definecolor{DarkRed}{rgb}{0.8,0,0}
\definecolor{Red}{rgb}{1,0,0}
\usepackage[linktocpage=true,
pagebackref=true,colorlinks,
linkcolor=DarkRed,citecolor=ForestGreen,
bookmarks,bookmarksopen,bookmarksnumbered]
{hyperref}

\usepackage[lined,boxed,commentsnumbered,ruled,vlined,noend,linesnumbered,boxed]{algorithm2e}

\usepackage{aliascnt}
\usepackage{hyperref}
\usepackage{cleveref}

\newtheorem{theorem}{Theorem}
\crefname{theorem}{theorem}{theorems}
\Crefname{theorem}{Theorem}{Theorems}

\newaliascnt{lemma}{theorem}
\newtheorem{lemma}[lemma]{Lemma}
\aliascntresetthe{lemma}
\crefname{lemma}{lemma}{lemmas}
\Crefname{lemma}{Lemma}{Lemmas}

\newaliascnt{proposition}{theorem}

\aliascntresetthe{proposition}
\crefname{proposition}{proposition}{propositions}
\Crefname{proposition}{Proposition}{Propositions}

\newaliascnt{corollary}{theorem}

\aliascntresetthe{corollary}
\crefname{corollary}{corollary}{corollaries}
\Crefname{corollary}{Corollary}{Corollaries}

\newaliascnt{claim}{theorem}

\aliascntresetthe{claim}
\crefname{claim}{claim}{claims}
\Crefname{claim}{Claim}{Claims}

\theoremstyle{definition}
\newaliascnt{definition}{theorem}
\newtheorem{definition}[definition]{Definition}
\aliascntresetthe{definition}
\crefname{definition}{definition}{definitions}
\Crefname{definition}{Definition}{Definitions}

\theoremstyle{remark}
\newaliascnt{remark}{theorem}
\newtheorem{remark}[remark]{Remark}
\aliascntresetthe{remark}
\crefname{remark}{remark}{remarks}
\Crefname{remark}{Remark}{Remarks}

 \theoremstyle{observation}
\newaliascnt{observation}{theorem}
\newtheorem{observation}[observation]{Observation}
\aliascntresetthe{observation}
\crefname{observation}{observation}{observations}
\Crefname{observation}{Observation}{Observations}

\newaliascnt{example}{theorem}

\aliascntresetthe{example}
\crefname{example}{example}{examples}
\Crefname{example}{Example}{Examples}

\usepackage{caption}
\usepackage{subcaption}

\usepackage{mathtools}

\newcommand{\defproblem}[3]{
  \vspace{1mm}
\noindent\fbox{
  \begin{minipage}{0.96\textwidth}
  \begin{tabular*}{\textwidth}{@{\extracolsep{\fill}}lr} #1  &  \\ \end{tabular*}
  {\bf{Input:}} #2  \\
  {\bf{Question:}} #3
  \end{minipage}
  }
  \vspace{1mm}
}
\usepackage{enumitem}

\definecolor{anti-flashwhite}{rgb}{0.95, 0.95, 0.96}

\usepackage{tabularx,lipsum,environ}
\usepackage{tikz}

\makeatletter
\newcommand{\problemtitle}[1]{\gdef\@problemtitle{#1}}
\newcommand{\probleminput}[1]{\gdef\@probleminput{#1}}
\newcommand{\problemquestion}[1]{\gdef\@problemquestion{#1}}
\NewEnviron{problem}{
	\problemtitle{}\probleminput{}\problemquestion{}
	\BODY
	\par\addvspace{.5\baselineskip}
	\noindent
	\begin{tabularx}{\textwidth}{@{\hspace{\parindent}} l X c}
		\multicolumn{2}{@{\hspace{\parindent}}l}{\@problemtitle} \\
		\textbf{Input:} & \@probleminput \\
		\textbf{Task:} & \@problemquestion
	\end{tabularx}
	\par\addvspace{.5\baselineskip}
}

\makeatletter
\providecommand*{\cupdot}{%
  \mathbin{%
    \mathpalette\@cupdot{}%
  }%
}
\newcommand*{\@cupdot}[2]{%
  \ooalign{%
    $\m@th#1\cup$\cr
    \hidewidth$\m@th#1\cdot$\hidewidth
  }%
}
\makeatother

\usepackage[lined,boxed,commentsnumbered,ruled,vlined,noend,linesnumbered,boxed]{algorithm2e}

 \usepackage{tcolorbox}

\usepackage{tikz}
\usetikzlibrary{calc,arrows.meta,decorations.pathreplacing}

\crefname{algocf}{Algorithm}{Algorithms}
\Crefname{algocf}{Algorithm}{Algorithms}
\newcommand{\Vis}{\mathrm{Vis}}

\newcommand{\cA}{\mathcal{A}}
\newcommand{\cK}{\mathcal{K}}
\newcommand{\cH}{\mathcal{H}}
\newcommand{\cT}{\mathcal{T}}

\newcommand{\cS}{\mathcal{S}}
\newcommand{\cQ}{\mathcal{Q}}
\newcommand{\mws}{\mu_{\mathrm{WS}}}

\newcommand{\nw}{\mathsf{nw}}

\tikzset{
  poly/.style={draw=black,thick,fill=gray!6},
  pt/.style={circle,fill=black,inner sep=1.3pt},
  win/.style={draw=red!75!black,very thick},
  visA/.style={fill=orange!35},
  visB/.style={fill=teal!30},
  visC/.style={fill=violet!25},
  gap/.style={fill=gray!45},
}

\newboolean{shortver}
\setboolean{shortver}{false}

\usepackage{graphicx} 

\usepackage{bm}
\usepackage{amsfonts}
\usepackage{tcolorbox}
\usepackage{mathtools}
\tcbuselibrary{skins}
\usepackage{xcolor}
\usepackage{cancel}
\usepackage{csquotes}
\usepackage[]{mdframed}
\usepackage{caption}
\usepackage{tcolorbox}
\tcbuselibrary{skins}
\usepackage[dvipsnames]{xcolor}
\usepackage{float}
\usepackage[bb=boondox]{mathalfa}
\usepackage{soul}
\usepackage{enumitem}

\colorlet{mix}{red!50!black}
\definecolor{DarkRed}{rgb}{0.8,0,0}
\definecolor{ForestGreen}{rgb}{0.1333,0.5451,0.1333}

\usepackage{hyperref}
\hypersetup{colorlinks={true},linkcolor={DarkRed},citecolor=ForestGreen}

\usepackage{tikz}
\usetikzlibrary{patterns,calc}

\usetikzlibrary{calc}

\usepackage{amsthm}
\usepackage{cleveref}

\newcommand{\vis}{\mathsf{Vis}}
\newcommand{\bd}{\mathsf{\delta}}

\newcommand{\inte}{\mathsf{iN}}

\newcommand{\pol}{\mathcal{P}}
\newcommand{\oo}{\mathcal{O}}

\usepackage[colorinlistoftodos]{todonotes}

\newcounter{sasanka}

\newcounter{satya}

\newtcolorbox{greenbox}[1][]{colback=green!60!white, colframe=green!50!black,
  boxrule=0.5pt, arc=2mm, left=2mm, right=2mm, top=1mm, bottom=1mm, #1}

\newcounter{debabrata}

\definecolor{bblue}{rgb}{0,0.1,0.55}

\definecolor{defblue}{rgb}{0.1,0.4,0.6} 

\newcommand{\sj}[1]{\todo[color=blue!40, fancyline]{#1}}
\newcommand{\deba}[1]{\todo[color=blue!40, fancyline]{#1}}

\begin{document}

\title{A Polynomial-Time Algorithm for Maximum Witness Set}

 \author{Satyabrata Jana  \thanks{Indian Institute of Science Education and Research Berhampur, Odisha, India, \textrm{satyamtma@gmail.com}}
 \and Debabrata Pal \thanks{Indian Statistical Institute, Kolkata, India,\textrm{ debabratapal4521@gmail.com}}
 \and Sasanka Roy \thanks{Indian Statistical Institute, Kolkata, India,\textrm{ sasanka.ro@gmail.com}}
}


 \maketitle

\thispagestyle{empty}

\begin{abstract}
A \emph{witness set} in a polygon is a set of points whose
visibility regions are pairwise disjoint. Since no point
of the polygon can see more than one witness, the
cardinality of a witness set provides a lower bound on
the minimum number of guards required to cover the
polygon. Finding a maximum witness set is therefore a
natural packing counterpart to the classical
\textsc{Art Gallery} problem. While the latter is
$\exists\mathbb{R}$-complete, the computational complexity
of finding a maximum witness set in a general simple
polygon has remained open. 


In this paper, we present a polynomial-time algorithm
for computing a maximum witness set in a simple polygon
with $n$ vertices, of which $\rho$ are reflex, assuming
that no three vertices are collinear. Our algorithm runs
in $\mathcal{O}(\rho^4n^6\log n)$ time. Our approach
exploits a tree structure induced by witness visibility regions and employs geometric dynamic
programming over nested regions determined by visibility windows.
\end{abstract}




\section{Introduction}\label{sec:intro}

Visibility-based optimization problems are among the central
topics in computational geometry. The classical
\textsc{Art Gallery} problem asks for a minimum set of guards
whose visibility regions together cover a given polygon \cite{ghosh2007visibility,DBLP:journals/siamrev/Franklin89}.
Two points $p,q$ in a polygon $\pol$ are said to be
\emph{visible} to each other if the line segment
$\overline{pq}$ lies entirely in $\pol$. The
\emph{visibility region} of a point $p\in\pol$, denoted
by $\Vis(p)$, consists of all points of $\pol$ visible
from $p$.

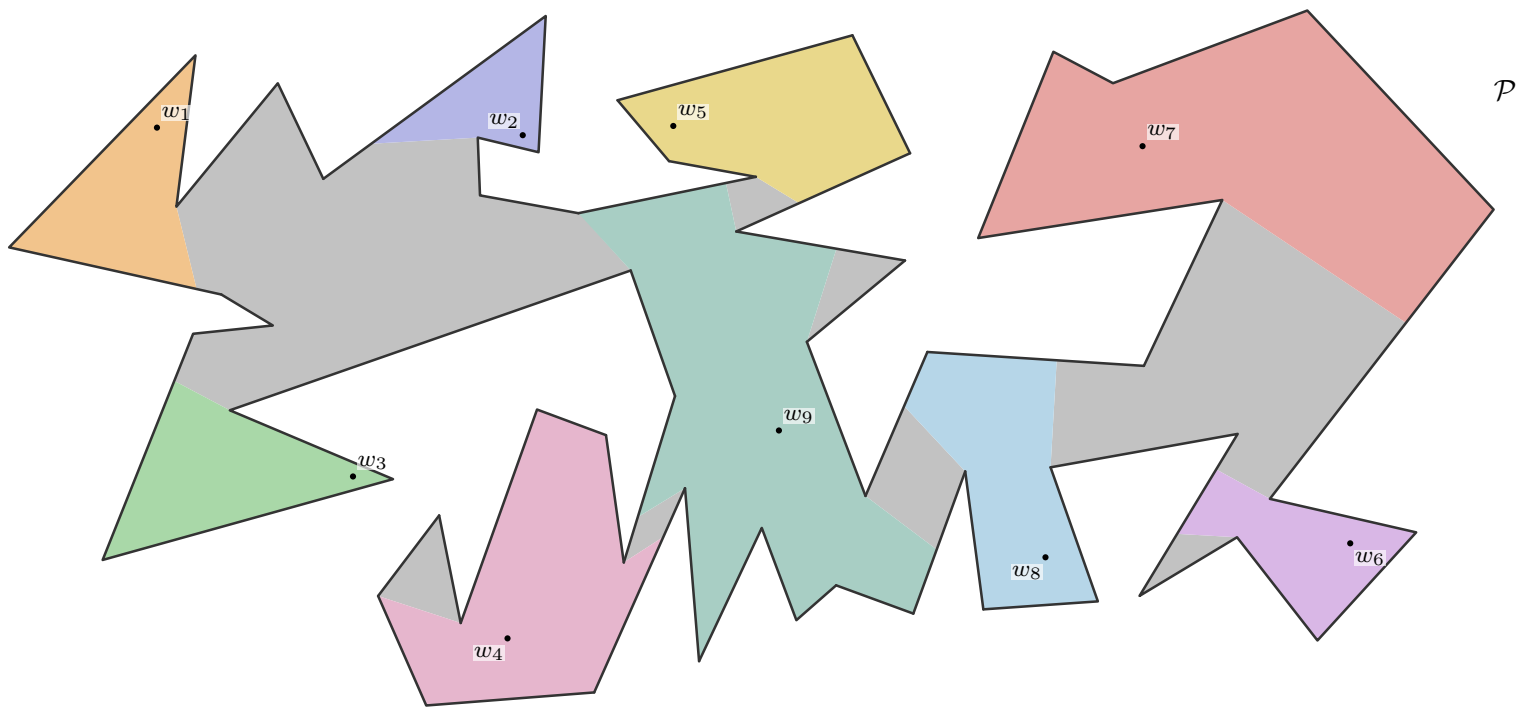
\begin{figure}[b!]
\centering

\begin{tikzpicture}[
    scale=0.55,
    transform shape,
    line cap=round,
    line join=round
]

\definecolor{visred}{HTML}{E7A5A2}
\definecolor{visgreen}{HTML}{A9D8A8}
\definecolor{visteal}{HTML}{A8CEC4}
\definecolor{visice}{HTML}{B6D6E8}
\definecolor{vispink}{HTML}{E6B6CD}
\definecolor{visgold}{HTML}{E9D88B}
\definecolor{visorange}{HTML}{F2C48C}
\definecolor{visviolet}{HTML}{B4B6E6}
\definecolor{vislavender}{HTML}{D9B7E6}

\begin{scope}[shift={(-0.35,-0.3)}]

\path[fill=black!24,draw=none]
(15.577,2.472) -- (14.857,2.513) --
(14.405,1.767) -- cycle;

\path[fill=black!24,draw=none]
(14.457,4.533) -- (15.400,6.528) --
(17.608,5.050) -- (15.972,2.934) --
(15.327,3.288) -- (15.585,3.715) --
(13.335,3.313) -- (13.409,4.600) --
cycle;

\path[fill=black!24,draw=none]
(10.761,5.942) -- (11.585,5.800) --
(10.405,4.822) -- cycle;

\path[fill=black!24,draw=none]
(6.448,7.278) -- (6.475,6.583) --
(7.668,6.372) -- (8.287,5.680) --
(3.467,3.999) -- (2.800,4.352) --
(3.026,4.919) -- (3.984,5.019) --
(3.365,5.391) -- (3.068,5.457) --
(2.825,6.447) -- (4.044,7.931) --
(4.591,6.782) -- (5.167,7.204) --
cycle;

\path[fill=black!24,draw=none]
(5.984,2.734) -- (6.243,1.441) --
(5.250,1.766) -- cycle;

\path[fill=black!24,draw=none]
(9.792,6.806) -- (10.312,6.489) --
(9.556,6.149) -- (9.432,6.731) --
cycle;

\path[fill=black!24,draw=none]
(8.372,2.714) -- (8.940,3.062) --
(8.680,2.477) -- (8.204,2.169) --
cycle;

\path[fill=black!24,draw=none]
(11.574,4.050) -- (12.309,3.264) --
(11.966,2.324) -- (11.110,2.969) --
cycle;


\path[fill=visred,draw=none]
(17.608,5.050) -- (18.662,6.413) --
(16.422,8.806) -- (14.086,7.932) --
(13.369,8.310) -- (12.465,6.070) --
(15.400,6.528) -- cycle;

\path[fill=visgreen,draw=none]
(2.800,4.352) -- (3.467,3.999) --
(5.429,3.171) -- (1.938,2.199) --
cycle;

\path[fill=visteal,draw=none]
(11.966,2.324) -- (11.685,1.554) --
(10.757,1.894) -- (10.279,1.475) --
(9.863,2.584) -- (9.109,0.982) --
(8.940,3.062) -- (8.372,2.714) --
(8.820,4.170) -- (8.287,5.680) --
(7.658,6.369) -- (9.432,6.731) --
(9.556,6.149) -- (10.761,5.942) --
(10.405,4.822) -- (11.110,2.969) --
cycle;

\path[fill=visice,draw=none]
(13.409,4.600) -- (13.335,3.313) --
(13.904,1.701) -- (12.527,1.605) --
(12.309,3.264) -- (11.574,4.050) --
(11.854,4.700) -- cycle;

\path[fill=vispink,draw=none]
(8.680,2.477) -- (8.204,2.169) --
(7.990,3.700) -- (7.161,4.008) --
(6.243,1.441) -- (5.250,1.766) --
(5.829,0.449) -- (7.848,0.606) --
cycle;

\path[fill=visgold,draw=none]
(10.312,6.489) -- (11.646,7.090) --
(10.953,8.508) -- (8.129,7.727) --
(8.748,6.995) -- (9.792,6.806) --
cycle;

\path[fill=visorange,draw=none]
(3.068,5.457) -- (0.814,5.958) --
(3.054,8.267) -- (2.825,6.447) --
cycle;

\path[fill=visviolet,draw=none]
(5.167,7.204) -- (6.448,7.278) --
(7.179,7.102) -- (7.265,8.740) --
cycle;

\path[fill=vislavender,draw=none]
(15.327,3.288) -- (15.972,2.934) --
(17.731,2.531) -- (16.544,1.231) --
(15.577,2.472) -- (14.857,2.513) --
cycle;


\draw[black!80,line width=0.48pt]
(7.658,6.369) -- (6.475,6.583) --
(6.448,7.278) -- (7.179,7.102) --
(7.265,8.740) -- (4.591,6.782);

\draw[black!80,line width=0.48pt]
(12.309,3.264) -- (12.527,1.605);

\draw[black!80,line width=0.48pt]
(12.309,3.264) -- (11.685,1.554);

\draw[black!80,line width=0.48pt]
(11.854,4.700) -- (11.110,2.969);

\draw[black!80,line width=0.48pt]
(11.110,2.969) -- (10.405,4.822);

\draw[black!80,line width=0.48pt]
(8.940,3.062) -- (9.109,0.982);

\draw[black!80,line width=0.48pt]
(8.940,3.062) -- (7.848,0.606);

\draw[black!80,line width=0.48pt]
(8.820,4.170) -- (8.204,2.169);

\draw[black!80,line width=0.48pt]
(8.204,2.169) -- (7.990,3.700);

\draw[black!80,line width=0.48pt]
(7.990,3.700) -- (7.161,4.008) --
(6.243,1.441);

\draw[black!80,line width=0.48pt]
(6.243,1.441) -- (5.984,2.734);

\draw[black!80,line width=0.48pt]
(7.848,0.606) -- (5.829,0.449) --
(5.250,1.766);

\draw[black!80,line width=0.48pt]
(5.250,1.766) -- (5.984,2.734);

\draw[black!80,line width=0.48pt]
(11.854,4.700) -- (14.457,4.533);

\draw[black!80,line width=0.48pt]
(13.335,3.313) -- (13.904,1.701) --
(12.527,1.605);

\draw[black!80,line width=0.48pt]
(11.685,1.554) -- (10.757,1.894);

\draw[black!80,line width=0.48pt]
(10.757,1.894) -- (10.279,1.475) --
(9.863,2.584) -- (9.109,0.982);

\draw[black!80,line width=0.48pt]
(10.405,4.822) -- (11.585,5.800);

\draw[black!80,line width=0.48pt]
(11.585,5.800) -- (9.556,6.149);

\draw[black!80,line width=0.48pt]
(11.646,7.090) -- (9.556,6.149);

\draw[black!80,line width=0.48pt]
(9.792,6.806) -- (8.748,6.995);

\draw[black!80,line width=0.48pt]
(11.646,7.090) -- (10.953,8.508);

\draw[black!80,line width=0.48pt]
(9.792,6.806) -- (7.658,6.369);

\draw[black!80,line width=0.48pt]
(10.953,8.508) -- (8.129,7.727);

\draw[black!80,line width=0.48pt]
(8.129,7.727) -- (8.748,6.995);

\draw[black!80,line width=0.48pt]
(8.820,4.170) -- (8.287,5.680);

\draw[black!80,line width=0.48pt]
(8.287,5.680) -- (3.467,3.999) --
(5.429,3.171) -- (1.938,2.199) --
(3.026,4.919) -- (3.984,5.019) --
(3.365,5.391);

\draw[black!80,line width=0.48pt]
(4.591,6.782) -- (4.044,7.931) --
(2.825,6.447) -- (3.054,8.267) --
(0.814,5.958) -- (3.365,5.391);

\draw[black!80,line width=0.48pt]
(13.335,3.313) -- (15.585,3.715) --
(14.405,1.767) -- (15.577,2.472) --
(16.544,1.231) -- (17.731,2.531) --
(15.972,2.934) -- (18.662,6.413) --
(16.422,8.806) -- (14.086,7.932) --
(13.369,8.310) -- (12.465,6.070) --
(15.400,6.528) -- (14.457,4.533);


\fill[black] (2.591,7.398) circle (0.037);
\node[
    font=\scriptsize,
    fill=white,
    fill opacity=.65,
    text opacity=1,
    inner sep=.3pt
] at (2.831,7.568) {$w_{1}$};

\fill[black] (6.989,7.307) circle (0.037);
\node[
    font=\scriptsize,
    fill=white,
    fill opacity=.65,
    text opacity=1,
    inner sep=.3pt
] at (6.769,7.477) {$w_{2}$};

\fill[black] (4.948,3.203) circle (0.037);
\node[
    font=\scriptsize,
    fill=white,
    fill opacity=.65,
    text opacity=1,
    inner sep=.3pt
] at (5.188,3.373) {$w_{3}$};

\fill[black] (6.806,1.256) circle (0.037);
\node[
    font=\scriptsize,
    fill=white,
    fill opacity=.65,
    text opacity=1,
    inner sep=.3pt
] at (6.586,1.076) {$w_{4}$};

\fill[black] (8.799,7.418) circle (0.037);
\node[
    font=\scriptsize,
    fill=white,
    fill opacity=.65,
    text opacity=1,
    inner sep=.3pt
] at (9.039,7.588) {$w_{5}$};

\fill[black] (16.938,2.399) circle (0.037);
\node[
    font=\scriptsize,
    fill=white,
    fill opacity=.65,
    text opacity=1,
    inner sep=.3pt
] at (17.178,2.219) {$w_{6}$};

\fill[black] (14.441,7.175) circle (0.037);
\node[
    font=\scriptsize,
    fill=white,
    fill opacity=.65,
    text opacity=1,
    inner sep=.3pt
] at (14.681,7.345) {$w_{7}$};

\fill[black] (13.274,2.231) circle (0.037);
\node[
    font=\scriptsize,
    fill=white,
    fill opacity=.65,
    text opacity=1,
    inner sep=.3pt
] at (13.054,2.051) {$w_{8}$};

\fill[black] (10.069,3.756) circle (0.037);
\node[
    font=\scriptsize,
    fill=white,
    fill opacity=.65,
    text opacity=1,
    inner sep=.3pt
] at (10.309,3.926) {$w_{9}$};

\node[font=\small] at (18.8,7.85) {$\mathcal P$};

\end{scope}
\end{tikzpicture}

\caption{A simple polygon $\mathcal P$ with a witness set
$W=\{w_1,\ldots,w_9\}$. The colored regions are the
visibility regions of the witnesses; they are
pairwise disjoint.}

\label{fig:witness-nine}

\end{figure}

A natural packing counterpart to the \textsc{Art Gallery}
problem is the \textsc{Witness Set} problem, introduced
by Amit, Mitchell, and
Packer~\cite{DBLP:journals/ijcga/AmitMP10}.
A point set $W\subseteq\pol$ is called a
\emph{witness set} if the visibility regions of its points
are pairwise disjoint, that is,
$\Vis(u)\cap\Vis(v)=\emptyset$ for every pair of
distinct witnesses $u,v\in W$ (see
\Cref{fig:witness-nine} for an illustration).
Equivalently, no point of the polygon can see more than
one witness. Consequently, every guard set must contain
at least $|W|$ guards, and hence the maximum cardinality
of a witness set provides a lower bound on the minimum
number of guards required to cover the polygon. Such
lower bounds are particularly useful in exact art gallery
algorithms, where matching lower and upper bounds certify
the optimality of a guard configuration.

Despite this close connection, the two problems impose
fundamentally different geometric requirements. While
the \textsc{Art Gallery} problem seeks a minimum collection
of visibility regions whose union covers the polygon,
the \textsc{Witness Set} problem seeks a maximum
collection of pairwise disjoint visibility regions.
Moreover, the witness-set condition is stronger than
requiring the selected points to be mutually invisible.
Indeed, two witnesses must not only be invisible to
each other, but there must be no point anywhere in the
polygon that sees both. Thus, a witness set is also a
hidden set, but the converse need not hold.

In this paper, we study the following optimization
problem.

\begin{tcolorbox}[
    enhanced,
    title={\color{black}\textsc{Maximum Witness Set}},
    colback=white,
    boxrule=0.4pt,
    attach boxed title to top center={
        xshift=-5cm,yshift*=-2.5mm
    },
    boxed title style={
        size=small,
        frame hidden,
        colback=white
    }
]
\vspace{-1mm}
\textbf{Input:} A simple polygon $\pol$ with $n$ vertices.

\textbf{Output:} A maximum-cardinality point set
$W\subseteq\pol$ such that
$\Vis(u)\cap\Vis(v)=\emptyset$ for every pair of
distinct points $u,v\in W$.
\end{tcolorbox}


We denote the maximum cardinality of a witness set
in $\pol$ by $\mu_{\mathrm{WS}}(\pol)$.
Unlike many classical packing problems, the candidates
for a witness set are not given explicitly: witnesses
may be placed anywhere in the polygon. Thus, even
though the input polygon has finitely many vertices,
the solution space is continuous and contains infinitely
many possible witness configurations. The main challenge
is to identify enough geometric structure to avoid
searching this space explicitly.

\subsection{Motivation and Related Work}

The computational difficulty of geometric visibility
problems is closely related to the continuous nature
of their solution spaces. The \textsc{Art Gallery}
problem is known to be $\mathsf{NP}$-hard
~\cite{DBLP:journals/tit/LeeL86} and
$\mathsf{APX}$-hard
~\cite{DBLP:journals/algorithmica/EidenbenzSW01}.
In a major breakthrough, Abrahamsen, Adamaszek, and
Miltzow~\cite{AbrahamsenAM22} proved that the
\textsc{Art Gallery} problem is
$\exists\mathbb{R}$-complete, even for polygons with
integer coordinates. Consequently, it does not belong
to $\mathsf{NP}$ unless
$\mathsf{NP}=\exists\mathbb{R}$. In particular, this
rules out a polynomial-size certificate consisting
of polynomial-bit rational guard coordinates unless
these complexity classes coincide. 

The packing nature of the \textsc{Witness Set} problem
suggests a potentially different algorithmic behavior.
A witness set requires its visibility regions to be
mutually disjoint, imposing a strong separation
condition absent from the guarding problem. Although
this separation requirement makes the feasibility
condition more restrictive, it may also provide
additional geometric structure that can be exploited
algorithmically. In particular, one may hope to
decompose a witness set into smaller collections
whose visibility regions are separated by suitable
geometric boundaries.

Amit, Mitchell, and
Packer~\cite{DBLP:journals/ijcga/AmitMP10}
introduced visibility-independent witness points
in the context of computing lower bounds for polygon
guarding. They used large witness sets to assess
guarding heuristics and to certify optimality when
the resulting lower bounds matched the sizes of
guard sets. This established witness sets as a useful
tool in the study of exact guarding problems.
Nevertheless, the algorithmic problem of finding a
maximum witness set remained much less understood
than its covering counterpart.

Early exact results were obtained for special polygon
classes. Daescu, Friedrichs, Malik, Polishchuk, and
Schmidt~\cite{DBLP:journals/comgeo/DaescuFMPS19}
developed a linear-time algorithm for guarding
monotone mountains and showed that the minimum
guard number equals the maximum witness number
for this class. More recently, Das
et al.~\cite{DBLP:journals/corr/abs-2511-10224, Das2026Witness}
studied the continuous \textsc{Witness Set} problem
in monotone polygons. They obtained an exact
algorithm with running time
$\rho^{\mathcal{O}(k)}n^{\mathcal{O}(1)}$,
where $\rho$ is the number of reflex vertices and
$k$ is the desired witness-set size, together with
a polynomial-time approximation scheme.
They also gave a polynomial-time algorithm for the
discrete version of the problem in simple polygons.

A further development is due to Das, Mondal, and
Roy~\cite{DBLP:journals/corr/abs-2609-24460},
who obtained polynomial-time algorithms for both
the continuous and discrete versions of
\textsc{Witness Set} in weak visibility polygons.
Their continuous algorithm runs in
$\mathcal{O}(n\log n+\rho^2(n+\rho^2))$ time.
These results exploit structural properties of
weak visibility polygons, in which every point is
visible from some point of a fixed boundary edge.
However, arbitrary simple polygons need not possess
such a distinguished edge or the associated
geometric structure.

For general simple polygons, Jana
et al.~\cite{DBLP:journals/corr/abs-2605-01592}
established that the decision version of
\textsc{Witness Set} belongs to both
$\mathsf{NP}$ and $\mathsf{XP}$.
Their approach constructs a finite candidate set
of size $n^{f(k)}$, for a computable function
$f$, and reduces the continuous problem to a
discrete instance. This yields an exact algorithm
whose running time is polynomial for every fixed
value of $k$, but whose exponent depends on $k$.
The same work also established $\mathsf{NP}$-completeness
of the discrete variant for rectilinear polygons
with holes.

Thus, while the complexity of witness sets in
several restricted classes of polygons has been
investigated, the existence of a polynomial-time
algorithm for arbitrary simple polygons has remained
open. In particular, the parameter-dependent
approach for general simple polygons does not
resolve the problem when the size of an optimal
witness set is unrestricted.

This motivates the central question of our work:

\begin{quote}
\emph{Can the geometric separation of visibility
regions be exploited to compute a maximum witness
set in a general simple polygon in polynomial time?}
\end{quote}

Our approach answers this question in the real-RAM
model by exploiting a recursive structure induced
by witness sets. Rather than enumerating candidate
witness configurations, we represent subproblems
by suitably chosen geometric regions and combine
their solutions using dynamic programming.

\subsection{Our Results and Method}

Our main result is a polynomial-time algorithm
for computing a maximum witness set in a general
simple polygon. We assume that no three vertices
of the input polygon are collinear. Let $n$ be
the number of vertices of the polygon and $\rho$
the number of its reflex vertices.

\begin{theorem}\label{thm:main}
For a simple polygon $\pol$ with $n$ vertices,
$\rho$ of them reflex, and no three vertices
on a line, a maximum witness set can be computed
in $\mathcal{O}(\rho^4n^6\log n)$ time.
\end{theorem}

Our algorithm is based on a geometric dynamic
programming approach. Its main insight is that
the separation of visibility regions induces
a recursive structure, allowing a witness set
to be decomposed into smaller subproblems.
We next outline the geometric ingredients that
make this approach possible.

\medskip
\noindent
\emph{A tree structure induced by witness sets.}
Consider a witness set $W$ in a simple polygon
$\pol$. Since the visibility regions of distinct
witnesses are disjoint, the regions not visible
from any witness separate these visibility
regions. The boundaries of visibility regions
are partly determined by \emph{windows}, which
are segments extending from reflex vertices
to the polygon boundary.

We establish a structural property showing
that the visibility regions of the witnesses,
together with the intervening non-visible
components, induce a tree whose adjacencies
are determined by these windows. Rooting this
tree at a witness allows us to organize the
remaining witnesses into subtrees. Each such
subtree corresponds to a smaller geometric
subproblem, separated from the rest of the
configuration by an appropriate window.

\medskip
\noindent
\emph{Nested caps and compact subproblems.}
A window separates the polygon into two sides,
one of which contains the corresponding
witness. We call this side a \emph{cap}.
The visibility region of the witness lies
entirely within its cap, and a cap may
also contain the visibility regions of
several other witnesses belonging to
the same subtree.

The crucial geometric observation is that
caps associated with the same reflex vertex
and the same side are totally ordered by
inclusion. Thus, among all caps of a fixed
type supporting the same number of witnesses,
it is sufficient to retain an extremal
one. This nesting property provides the
state compression underlying our dynamic
programming algorithm.

\medskip
\noindent
\emph{Dynamic programming over caps.}
We associate a state with a triple
$(\alpha,r,i)$, where $\alpha$ denotes
the number of witnesses, $r$ is a reflex
vertex, and $i\in\{1,2\}$ specifies one
of the two sides associated with $r$.
The corresponding dynamic programming
entry represents the smallest limiting
cap of type $(r,i)$ that can support
$\alpha$ witnesses.

By the classical art gallery theorem,
every witness set contains at most
$\lfloor n/3\rfloor$ points.
Consequently, the number of states
is at most
$2\rho\lfloor n/3\rfloor$,
which is polynomial in the input size.

To compute a state, we select a root
witness and combine solutions to
smaller subproblems. The caps
containing these subproblems must
be pairwise disjoint, lie inside
the cap associated with the root,
and avoid the root's visibility
region. We show that the relevant
compatibility conditions can be
expressed through boundary arcs
of caps. This reduces the selection
of compatible child subproblems
to a weighted interval scheduling
problem, with weights corresponding
to the numbers of witnesses
supported by the child caps.

\medskip
\noindent
\emph{Handling the continuous search space.}
A remaining difficulty is that a
root witness can be placed anywhere
inside the polygon. Consequently,
even though the dynamic programming
table has polynomially many states,
each state initially involves
infinitely many possible witness
locations.

We overcome this difficulty by
constructing a suitable arrangement
of lines. The arrangement is
determined by the polygon edges,
lines through reflex vertices,
and geometric regions associated
with previously computed states.
These lines partition the polygon
into polynomially many cells.

The key property is that, within
an open arrangement cell, the
combinatorial information relevant
to the dynamic programming
recurrence remains unchanged.
In particular, the available
window types and the admissibility
of previously computed subproblems
are constant throughout the cell.
Thus, a single representative
point suffices to determine the
feasibility of a state within
that cell.

Moreover, the nesting of caps
allows us to determine an extremal
cap for each feasible cell by
considering an appropriate extreme
corner. Processing the states in
increasing order of the number
of witnesses yields the claimed
polynomial-time algorithm.

\medskip
\noindent
\emph{Computing an optimal witness set.}
After the dynamic programming table
has been constructed, the algorithm
considers an unrestricted root witness
and combines compatible previously
computed subproblems. A final
weighted interval scheduling
computation determines the maximum
number of witnesses that can be
supported.

The extremal objects stored in
the table may arise as limits
of ordinary caps. The algorithm
therefore records the geometric
information needed to choose
appropriate witness positions
and recover an optimal solution.

\medskip
The main advantage of our approach
is that it avoids enumerating
witness configurations. Instead,
the tree structure of witness
visibility regions provides a
recursive decomposition, the
nesting of caps permits a compact
representation of subproblems,
and line arrangements together
with weighted interval scheduling
allow the resulting dynamic
programming recurrence to be
evaluated efficiently. Our running-time bound is stated
in the real-RAM model,  in which arithmetic operations and comparisons involving real numbers are assumed to take constant time.

\paragraph*{Organization.}
\Cref{sec:prelim} introduces the
necessary geometric definitions
and notation. We then establish
the tree structure induced by
witness sets and the properties
of windows and caps.
In \cref{sec:table}, we develop
the dynamic programming formulation.
Finally, \cref{sec:alg} presents
the algorithm, proves its
correctness, and analyzes
its running time.

\section{Preliminaries}\label{sec:prelim}

\begin{definition}
{\em A \emph{polygon} is a closed region in the plane bounded by a finite sequence of non-intersecting line segments, called its \emph{edges}. The endpoints of these edges are its \emph{vertices}. A \emph{polygon with holes} is a polygon whose boundary consists of more than one disjoint simple cycle: one outer boundary and one or more interior boundaries.  A polygon is \emph{simple} if it is neither self-intersecting nor contains holes.}
\end{definition}
For a polygon $\pol$, we denote its boundary by \emph{$\bd(\pol)$} and its interior by \emph{$\inte(\pol)$}. A vertex of $\pol$ is \emph{convex} if its interior angle is at most $180^\circ$; otherwise, it is \emph{reflex}. Let $n$ and $\rho$ be the number of vertices and the reflex vertices of $\pol$ respectively. For a line segment $[a,b]$, we denote its interior by \emph{$iN([a,b])$} $\coloneq[a,b]\setminus\{a,b\}$ and its boundary by \emph{$\bd([a,b])$} $\coloneq \{a,b\}$.
\begin{definition}
{\em Two points $x,y\in \pol$ are \emph{visible} to each other if the segment $\overline{xy}$ is contained entirely in $\pol$. The \emph{visibility region} of $p$ is $\vis(p)=\{q\in \pol : \overline{pq}\subseteq \pol\}$. For a set $S\subseteq \pol$, $\vis(S)=\bigcup_{p\in S}\vis(p)$.}
\end{definition}

\begin{definition}
{\em In a polygon $\pol$, a finite point set $W\subseteq \pol$ is said to be a \emph{witness set} of $\pol$ if $\vis(x)\cap\vis(y)=\emptyset\text{ for all distinct } x,y\in W$. A witness point is a point of $W$. See \Cref{fig:witness-nine}.}
\end{definition}


\begin{lemma}\label{lem:size}
Every witness set has at most $\lfloor n/3\rfloor$ points.
\end{lemma}

\begin{proof}
By Fisk's proof of the art gallery theorem~\cite{DBLP:journals/jct/Fisk78a}, some $\lfloor n/3\rfloor$ vertices see all of $\pol$. Each of them sees at most one witness.
\end{proof}


\begin{definition}\label{window}
    {\em Let $p\in \pol$. Suppose $p$ sees a reflex vertex $r$, and the ray from $p$ through $r$ goes on into the interior of $\pol$ after $r$. Define \emph{$h_r(p)$}$~\in \bd(\pol)$ to be the boundary point on this ray such that it goes outside $\pol$ after $h_r(p)$. The segment $F=[r,h_r(p)]$ is a \emph{window} of $p$.}
\end{definition}

For a point $p\in \pol$, let $F=[r,h_r(p)]$ be a window of $p$ and $e_1, e_2$ be the two incident edges of $r$. For $i\in \{1,2\}$, let $\gamma_i$ be the path from $r$ to $h_r(p)$ along $\bd(\pol)$ such that $e_i$ lies in $\gamma_i$. For $i\in \{1,2\}$, let $R_i$ be the region bounded by the path $\gamma_i$ and the line segment $[r,h_r(p)]$. Then, $R_1\cup R_2=\pol$ (since $\gamma_1\cup\gamma_2=\bd(\pol)$) and exactly one of $R_i$ contains $p$.
\begin{definition}\label{pocketsdd}
    {\em  We call $R_i$ the \emph{contained side} of $F$ if it contains $p$, written \emph{$P_F(p)$}, and the other region the \emph{pocket side} of $F$.} 
\end{definition}
\begin{definition}
    {\em  A point $p$ has \emph{type} $(r,i)$, for $i\in\{1,2\}$, if $R_i$ contains $p$. We define the \emph{cap} of $p$ associated with $(r,i)$ as $P_F(p)$ and denote it by \emph{$H_{r,i}(p)$}. We define its \emph{boundary arc $A(H_{r,i}(p))$}$\,\coloneq H_{r,i}(p)\cap\bd(\pol)$, which is precisely $\gamma_i$. We also define the \emph{corresponding window} of $H_{r,i}(p)$ to be \emph{$F(H_{r,i}(p))$}$\,\coloneq F$.}
\end{definition}
\begin{remark}
    There are cases where both the pocket side and the contained side of a window of a point is not simple polygons. This happens when $p$ lies on the line $l$ joining two reflex vertices (both of which are visible from $p$ and lies on the same side of $p$) such that the pairs of edges of those reflex vertices lies on the either side of $l$. Note that, in that case, $\vis(p)$ is also not a simple polygon. See page 4-5 of \cite{DBLP:journals/corr/abs-2605-01592} for detailed notion.
\end{remark} 

The following observation is immediate from the definition of a window.
\begin{observation}\label{lem:confine}
For every window $F$ of a point $w$, we have $\Vis(w)\subseteq P_F(w)$.
\end{observation}


\begin{definition}
    {\em For two points $p,q\in \pol$, if $q\notin\Vis(p)$, then $q$ lies in exactly one pocket side of $p$ for some window of $p$. We call it a window of $p$ \emph{facing} $q$ and denote it by $C_p(q)$.}
\end{definition}

\begin{lemma}\label{lem:facing}
Let $p, q\in\pol$ with $q\notin\Vis(p)$. Put $C=C_p(q)$ and $D=C_q(p)$. Then
$\Vis(p)\cap\Vis(q)=\emptyset$ if and only if $C$ and $D$ are disjoint and $D$ lies on the pocket
side of $C$.
\end{lemma}

\begin{proof}
If the regions are disjoint, so are $C\subseteq\Vis(p)$ and $D\subseteq\Vis(q)$. The region $\Vis(q)$ is connected, contains $q$ and $D$, and misses $C$, so it stays on the pocket side of $C$. Conversely, two disjoint chords cut $\pol$ into two outer parts and a middle part. The hypotheses place $p$ and $q$ in different outer parts, and \Cref{lem:confine} places their visibility regions there as well.
\end{proof}

\section{Witness Tree}
\begin{definition}
    {\em For a witness set $W$, define the graph \emph{$\cT(W)$} with one node for each region $\Vis(w)$, $w\in W$, and one node for each connected component of $\pol\setminus\bigcup_w\Vis(w)$. Join two nodes if their regions meet along a window.}
\end{definition}

\begin{remark}
\begin{itemize}[
    leftmargin=1.5em,
    labelsep=0.5em,
    itemsep=2pt,
    topsep=2pt,
    parsep=0pt
]
    \item For any two witnesses $w_1,w_2\in W$,
    their distance in $\mathcal{T}(W)$ is always even,
    since there is at least one connected component of
    $P\setminus\bigcup_{w\in W}\Vis(w)$ between
    $\Vis(w_1)$ and $\Vis(w_2)$.

    \item For a node corresponding to $\Vis(w)$,
    we simply use $w$ as shorthand notation.
\end{itemize}
\end{remark}

     

\begin{lemma}\label{thm:tree}
For a witness set $W$, $\cT(W)$ is a tree.
\end{lemma}

\begin{proof}
    For each $w\in W$, the windows of $w$ lies in distinct rays from $w$ and no window contains $w$. So all the windows of $w$ are disjoint. Windows of different witnesses lie in disjoint visibility regions, so they are disjoint. Let $S=\{s_1,\ldots, s_d\}$ denote the collection of all windows of the points in $W$. We cut $\pol$ along these windows one at a time. For each $i\in[d]$, we consider a graph $G_i$ with one node for each connected component of $\pol\setminus \bigcup_{j\leq i}s_j$ and there is an edge between two nodes if their corresponding components lie on either side of some $s_j,~ j\leq i$. Observe that, $G_d=\cT(W)$. Since $s_1$ is a chord, it divides the polygon into two parts, $G_1$ has two nodes which are connected by a single edge. So, $G_1$ is a tree. Now, assume that $G_{i-1}$ is a tree. Let $s_i$ lies on the component $c$ of $\pol\setminus\bigcup_{j\leq i-1}s_j$. $s_i$ cuts $c$ into two parts $c_1$ and $c_2$, so there is an edge between $c_1$ and $c_2$ in $G_i$ and hence $G_i$ is connected. Only one new edge has been introduced to $G_i$ and two new nodes replaced one old node. Thus $|E(G_i)|=|V(G_i)|-1$ since $|E(G_i)|=|E(G_{i-1})|+1$, $|V(G_i)|=|V(G_{i-1})|+1$ and $|E(G_{i-1})|=|V(G_{i-1})|-1$. So $G_i$ is a tree and hence $\cT(W)=G_d$ is a tree by induction.
\end{proof}



    

\begin{lemma}\label{lem:order}
    Let $r$ be a reflex vertex with incident edges $e_1$ and $e_2$. Let $C$ be the set of all caps associated with $(r,i)$ for $i\in \{1,2\}$, arising from the points of type $(r,i)$. Then the elements of $C$ are totally ordered by inclusion.

\end{lemma}

\begin{proof}
Two windows that start at $r$ cannot cross elsewhere. The cap with the shorter boundary arc is
therefore contained in the other one.
\end{proof}

\begin{observation}\label{windowinter}
    Let $K$ be a cap and $p,q$ be two points in $\pol$ such that $p\in K$, $q\notin K$. Then any path between $p$ and $q$ must intersect $F(K)$.
\end{observation}

\begin{lemma}\label{lem:arc}
    Let $K_1,K_2$ be two caps. Then the followings are true.
    \begin{enumerate}
        \item $K_1\cap K_2=\emptyset$ if and only if $A({K_1})\cap A({K_2})=\emptyset$.
        \item $K_1\subseteq K_2$ if and only if $A(K_1)\subseteq A(K_2)$.
    \end{enumerate}

\end{lemma}
\begin{proof}
    For (1), one direction holds because the arcs lie in the caps. For the reverse direction, observe that $K_j$ is precisely the closed region bounded by $A_{K_j}$ and $F_{K_j}$. Since the endpoints of $F_{K_j}$ lies in $A_{K_j}$ and $A_{K_1}\cap A_{K_2}=\emptyset$, $F_{K_1}$ and $F_{K_2}$ are also disjoint. Thus the regions bounded by $A_{K_j}$ and $F_{K_j}$ are disjoint. Hence $K_1\cap K_2=\emptyset$.

    Now, let $K_1\subseteq K_2$. Then, $A(K_1)=K_1\cap \bd(\pol)\subseteq K_2\cap\bd(\pol)=A(K_2)$. For the reverse direction, let $A(K_1)\subseteq A(K_2)$. Since $K_i$ is the region bounded by the simple polygonal chain $A(K_i)$ and the line segment $F(K_i)$, $K_1\subseteq K_2$.
\end{proof}

\begin{definition}
    {\em For a cap $K$, define \emph{$R_K$} $\coloneq K\cup\Vis(F(K))$.}
\end{definition}

\begin{lemma}\label{lem:avoid}
    For a point $p\in \pol$, we have $\vis(p)\cap K=\emptyset$ if and only if $p\notin R_K$.
\end{lemma}
\begin{proof}
    Let $\vis(p)\cap K=\emptyset$. As $p\in \vis(p)$, $p\notin K$. If $p\in \vis(F_K)$, then there exists $q\in F_K\subseteq K$ such that $q$ sees $p$ which implies that $q\in K\cap\vis(p)$, a contradiction. Thus $p\notin R_K$. Conversely, let $x\in \vis(p)\cap K$. Now $p\notin K$ since $p\notin R_K$. So, using \Cref{windowinter}, the visibility line segment $[p,x]$ intersects $F_K$ at $y$ (let). Thus $p\in \vis(F_K)$, a contradiction.
\end{proof}
\begin{definition}
    {\em A witness set $W$ is \emph{confined} to a set $S$ if $\vis(w)\subseteq S$ for all $w\in W$}
\end{definition}

\begin{lemma}\label{lem:subtree}
    Root $\cT(W)$ at a witness point $w\in W$. For any other witness $u\in W$, let $F_u=C_u(w)$ (a window of $u$ facing $w$) and let $H_u=P_{F_u}(u)$. Let $W_u$ be the set of witness points which are in the subtree rooted at $u$. Then $W_u$ is confined to the cap $H_u$. Let $v_1,v_2\in W$ are distance two descendants of $u$ in $\cT(W)$. Then $H_{v_1}\cap H_{v_2}=\emptyset$, $H_{v_i}\subsetneq H_u$, and $H_{v_i}\cap\vis(u)=\emptyset$ for each $i\in \{1,2\}$.
      
    Moreover, if $W$ is confined to a cap $H$ of the root $w$, then $H_u\subseteq H$. 
\end{lemma}

\begin{proof}
    The chord $F_u$ divides $\pol$ into two pieces one of which is $H_u$ and the other one is the pocket side of $F_u$. Let $v\in W_u$. Clearly, $\vis(v)$ is contained in exactly one of the pieces since $\vis(v)\cap\vis(u)=\emptyset$, $F_u\subseteq\vis(u)$ and $\vis(v)$ is a connected region. Now, if $\vis(v)$ is contained in the pocket side of $F_u$, then $\vis(w)$ and $\vis(v)$ are in the same side of $F_u$. which implies there is a path in $\cT(w)$ from the root to $v$ which does not contain $u$. This contradicts the tree structure of $\cT(w)$ since $v$ lies in the subtree rooted at $u$ so the path from $w$ to $v$ must go through $u$. Thus, $\vis(v)\subset H_u$ and $W_u$ is confined to $H_u$. Observe that, $H_{v_i}$ is the union of the regions descendant to $v_i$ and $\vis(v_i)$. Thus $H_{v_1}\cap H_{v_2}=\emptyset$, since descendants of $v_1$ and $v_2$ are disjoint and $\vis(v_1)\cap \vis(v_2)= \emptyset$. Now, $\vis(u)$ lies on the pocket side of $F_{v_i}$ since $C_{v_i}(u)=C_{v_i}(w)$. So, $H_{v_i}\subsetneq H_u$ and $H_{v_i}\cap\vis(u)=\emptyset$. If $W$ is confined to a cap $H$ of $w$, then $H$ is the union of the regions descendant to $w$ and $\vis(w)$. Hence, $H_u\subseteq H$ because $H_u$ is also the union of the regions descendant to $u$ and $\vis(u)$.
\end{proof}
\begin{observation}\label{obs:realize}
    Let $w$ be a point and $H$ be a cap of $w$. For $i\in \{1,\ldots,m\}$, let $K_i$ be a cap such that the $K_i$ are pairwise disjoint, $K_i\cap \vis(w)=\emptyset$ and there is a witness set $W_i$ of size $t_i$ confined to 
    $K_i$. Then $W'=\{w\}\cup\left(\bigcup_{i=1}^mW_i\right)$ is a witness set of size $1+\sum_{i=1}^mt_i$. Moreover, if each $K_i$ is contained in a cap $H$ of $w$, then $W'$ is confined to $H$.
\end{observation}

\section{The Dynamic Programming}\label{sec:table}

\begin{definition}\label{def:state}
    {\em A \emph{state} is a triple $s=(\alpha,r,i)$ with $1\le \alpha\le \lfloor n/3\rfloor$, a reflex vertex $r$ and $i\in\{1,2\}$. A cap $H_{r,i}(w)$ of $w$ \emph{works} for $s$ if some witness set $U\subseteq R$ of size $\alpha$ contains $w$ and is confined to the cap. For a state $s=(\alpha, r,i)$, we define \emph{$\mathsf{nw}(s)$} $=\alpha$. Let \emph{$\cH(s)$} be the family of caps that work for $s$ and define \emph{$K^*(s)$}$\,\coloneq\bigcap_{H\in\cH(s)}H$.}
\end{definition}
    

Let $h=\lfloor n/3\rfloor$. There are at most $2\rho h$ states since $1\le \alpha\le h$, $\rho$ is the number of reflex vertices and $i\in \{1,2\}$. Deleting a witness keeps a set confined. Hence, we have the following observation.
\begin{observation}\label{eq:mono}
Let $\beta\le \alpha$. If a cap $H$ works for $(\alpha,r,i)$, then $H$ also works for $(\beta,r,i)$. Moreover, $K^*(\alpha,r,i)\neq\emptyset \implies K^*(\beta,r,i)\neq\emptyset$ and $K^*(\beta,r,i)\subseteq K^*(\alpha,r,i)$.
\end{observation}


\begin{definition}\label{def:adm}
{\em Let $w$ be regular with a cap $H$. A cap $K$ is called \emph{admissible for $w$ in $H$} if $\Vis(w)\cap K=\emptyset$ and $K\subseteq H$. $K$ is called \emph{root-admissible for $w$} if $\vis(w)\cap K=\emptyset$.}
\end{definition}

\begin{lemma}\label{lem:contain}
    Let $H$ be a cap of $w$, and let $K$ be a cap at some reflex vertex $r$ such that $\vis(w)\cap K=\emptyset$. Then $K\subseteq H$ if and only if $r\in A(H)$. 
\end{lemma}

\begin{proof}
    We have $\vis(w)\cap K=\emptyset$. For the forward direction, let $K\subseteq H$. Then, $r\in A(H)$, since $A(K)\subseteq A(H)$ (using \Cref{lem:arc}) and $r\in A(K)$. Now, for the reverse direction, let $r\in A(H)$. If $K$ is not contained in $H$, there exists $y\in K\setminus H$. Since $K$ is connected, there is a path $\gamma\subseteq K$ between $y$ and $r$. Now, $r\in H$ and $y\notin H$, so $\gamma$ must intersect $F(H)\subseteq \vis(w)$ using \Cref{windowinter}. Thus, $\vis(w)\cap K\neq \emptyset$, which is a contradiction. 
\end{proof}

\begin{lemma}\label{thm:rec}
    The cap $H_{r,i}(w)$ works for $s=(\alpha,r,i)$ if and only if there exist states $s_1,\ldots,s_m$ satisfying the following properties for some $m\ge 0$.
    \begin{enumerate}
        \item $\nw(s_j)<\nw(s)$, for each $j\in [m]$.
        \item $K^*(s_j)$ are pairwise disjoint and each is non-empty and admissible for $w$ in $H_{r,i}(w)$.
        \item $1+\sum_{j=1}^{m}\nw(s_j)\ge \alpha$.
    \end{enumerate}
    Clearly, for $\alpha=1$, the empty family is allowed since any cap of $w$ associated to $(r,i)$ works for $(1,r,i)$. 
\end{lemma}

\begin{proof}
    For the forward direction, let $H=H_{r,i}(w)$ works for $s=(\alpha,r,i)$. Let $W_w$ be a witness set confined to $H$ such that $w\in W_w$ and $|W_w|=\alpha$. Now, we consider the tree $\cT(W_w)$ rooted at $w$. Let $v_1,\ldots, v_m$ are the distance two descendants of $w$. Let $F_j=(C_{v_j}(w))$ and $H_j=P_{F_j}(v_j)$. Let $\alpha_j$ be the size of subtree of $\cT(W_w)$ rooted at $v_j$, and let $r_j$ be a reflex vertex and $i_j\in \{1,2\}$ be such that $H_j=H_{r_j,i_j}(v_j)$. So, $\alpha=1+\sum_{j=1}^m\alpha_j$. Define $s_j=(\alpha_j,r_j,i_j)$ for each $j\in[m]$. The conditions (1) and (3) are satisfied, since $\nw(s_j)=\alpha _j$. Now, using \Cref{lem:subtree}, $H_j$ are pairwise disjoint, $H_j\subsetneq H$, and $H_j\cap \vis(w)=\emptyset$. Thus, condition (2) is also satisfied because $K^*(s_j)\subseteq H_j$ for each $j$.

    For the reverse direction, let $s=(\alpha,r,i)$ be a state and there are states $s_1,\ldots, s_m$ satisfying the three conditions. Let $W_j$ be a witness set of size $\nw(s_j)$ confined in $K^*(s_j)$, and $W'=\{w\}\cup\left(\bigcup_{j=1}^mW_j\right)$, $\alpha'=1+\sum_{j=1}^m\nw(s_j)$. Thus, using \Cref{obs:realize}, $W'$ is a witness set of size $\alpha'$ confined in $H_{r,i}(w)$. So, $H_{r,i}(w)$ works for $(\alpha',r,i)$ as well as $(\alpha,r,i)$ using \Cref{eq:mono}.

\end{proof}
\begin{remark}\label{l1}
    Suppose $H_{r,i}(w)$ works for $s=(\alpha,r,i)$ then there exist states $s_1,\ldots,s_m$ satisfying the three properties of \Cref{thm:rec}. Now, by \Cref{eq:mono}, we can always choose a family $s_1',\ldots,s_m'$ such that it satisfies the first two properties of \Cref{thm:rec} with the equality $1+\sum_{j=1}^{m}\nw(s_j')=\alpha$. We call such a family a \emph{child record} of $s$.
\end{remark}
\section{Algorithm for Finding Maximum Witness Set}\label{sec:alg}
Define \emph{$\cK_{<\alpha}$} $\coloneq \{K~|~K=K^*(s)\text{ for some state $s$ with }\nw(s)<\alpha\}$. Let \emph{$L_1$} consists of the supporting lines of the edges of $\pol$ and \emph{$L_2$} be the set of lines through a reflex vertex and another vertex. Now, define \emph{$L_3^{(\alpha)}$} to be the set of supporting lines of the edges of $R_K$ for each $K\in \cK_{<\alpha}$.

\begin{definition}
    {\em The \emph{arrangement} of row $\alpha$, denoted by \emph{$\cA_\alpha$}, is defined as $L_1\cup L_2\cup L_3^{(\alpha)}$. The lines of $\cA_\alpha$ divide $\pol$ into convex regions, which we call \emph{cells}. Let \emph{$C_\alpha$} denote the set of all such cells contained in $\pol$. The intersection of two distinct cells, if nonempty, lies entirely on their boundaries. For a cell $c$, let \emph{$g(c)$} be the arithmetic mean of all vertices of $c$.}
\end{definition}

We have $|L_1|=\oo(n),~|L_2|=\oo(\rho n),~|L_3^{(\alpha)}|=\oo(\rho n^2)$ since there are at most $2\rho h=\oo(\rho n)$ states and $R_K$ has $\oo(n)$ edges for $K\in \cK_{<\alpha}$. So, there are $\oo(\rho n^2)$ lines in $\cA_\alpha$. Now, the number of cells created by the lines in $\cA_\alpha$ is $\oo(|\cA_\alpha|^2)$. Thus, $|C_\alpha|=\oo(\rho^2n^4)$. 

\begin{lemma}\label{lem:const}
    Let $c\in C_\alpha$ and $v_1,v_2$ be two points in $iN(c)$. Then the followings are true.
    \begin{enumerate}
        \item $v_1$ has type $(r,i)$ if and only if $v_2$ has type $(r,i)$, i.e., the available types inside a cell remains same.
        \item For an available type $(r,i)$ and for $K\in \cK_{<\alpha}$, $K$ is admissible for $v_1$ in $H_{r,i}(v_1)$ if and only if $K$ is admissible for $v_2$ in $H_{r,i}(v_2)$.
    \end{enumerate}

\end{lemma}

\begin{proof}
    (1) Let $v_1$ has type $(r,i)$. If $r$ is not visible from $v_2$, then there must be another reflex vertex $r'$ such that its incident edges blocks the visibility of $v_2$ and restrains $v_2$ from seeing $r$. So, the triangle $\triangle v_1v_2r$ contains $r'$. Thus , $v_1$ and $v_2$ lie on the either side of the line $l$ through $r$ and $r'$ (see \Cref{pf12}), which is a contradiction since $l\in L_2$ and $v_1,v_2$ are contained in the same cell $c$. Hence, $v_2$ sees $r$. Since $v_1, v_2$ are in the same cell, they lie on the same side of the supporting line of each incident edge of $r$ (see \Cref{pf121}). Thus, $v_2$ also has type $(r,i)$. The reverse direction is symmetric to the forward direction.

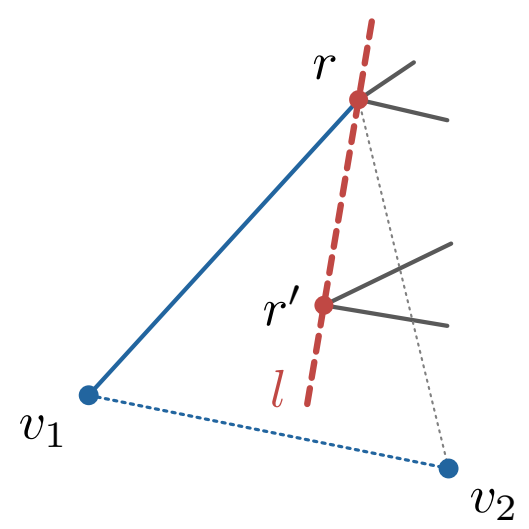
\begin{figure}[ht!]
    \centering
\begin{tikzpicture}[scale=0.85,line cap=round,line join=round,
    every node/.style={font=\small}]
  \definecolor{proofblue}{RGB}{34,101,159}
  \definecolor{proofred}{RGB}{192,72,68}
  \coordinate (r) at (0.15,2.55);
  \coordinate (rp) at (-0.12,0.95);
  \coordinate (vone) at (-1.95,0.25);
  \coordinate (vtwo) at (0.85,-0.32);
  \draw[proofred,densely dashed,very thick] ($(rp)!-0.48!(r)$) -- ($(rp)!1.43!(r)$)
    node[pos=0.04,left=0.1pt,text=proofred] {$l$};
  \draw[black!65,thick] (r) -- ++(0.69,-0.16);
  \draw[black!65,thick] (r) -- ++(0.43,0.29);
  \draw[black!65,thick] (rp) -- ++(0.99,0.48);
  \draw[black!65,thick] (rp) -- ++(0.96,-0.16);
  \draw[proofblue,thick] (vone) -- (r);
  \draw[proofblue,densely dotted,semithick] (vone) -- (vtwo);
  \draw[black!50,densely dotted] (vtwo) -- (r);
  \foreach \q in {r,rp} \fill[proofred] (\q) circle (2.1pt);
  \foreach \q in {vone,vtwo} \fill[proofblue] (\q) circle (2.2pt);
  \node[above left=0.1pt] at (r) {$r$};
  \node[left=0.1pt] at (rp) {$r'$};
  \node[below left=0.01pt] at (vone) {$v_1$};
  \node[below right=0.1pt] at (vtwo) {$v_2$};
\end{tikzpicture}
    \caption{The line through two reflex vertices
    separates $v_1$ and $v_2$.}
    \label{pf12}
\end{figure}

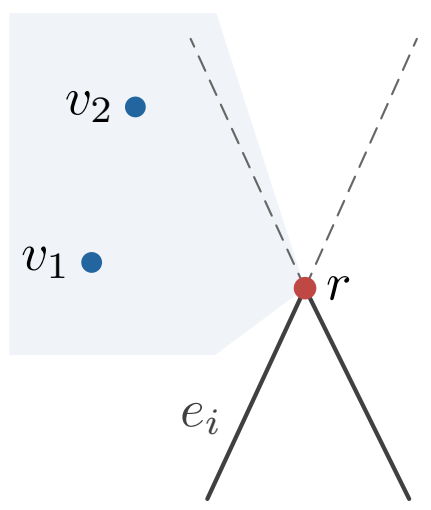
\begin{figure}[ht!]
    \centering
\begin{tikzpicture}[scale=0.85,line cap=round,line join=round,
    every node/.style={font=\small}]
  \definecolor{proofblue}{RGB}{34,101,159}
  \definecolor{proofred}{RGB}{192,72,68}
  \coordinate (r) at (0,0);
  \coordinate (vone) at (-1.66,0.20);
  \coordinate (vtwo) at (-1.32,1.41);
  \fill[proofblue!7] (r) -- (-0.69,2.14) -- (-2.30,2.14)
     -- (-2.30,-0.52) -- (-0.70,-0.52) -- cycle;
  \draw[black!75,thick] (r) -- (-0.76,-1.64)
    node[pos=0.62,left=0.1pt] {$e_i$};
  \draw[black!75,thick] (r) -- (0.81,-1.64);
  \draw[black!60,densely dashed] (r) -- (0.87,1.94);
  \draw[black!60,densely dashed] (r) -- (-0.89,1.94);
  \fill[proofblue] (vone) circle (2.3pt);
  \fill[proofblue] (vtwo) circle (2.3pt);
  \fill[proofred] (r) circle (2.5pt);
  \node[left=0.1pt] at (vone) {$v_1$};
  \node[left=0.1pt] at (vtwo) {$v_2$};
  \node[right=0.1pt] at (r) {$r$};
\end{tikzpicture}
    \caption{The positions of $v_1$ and $v_2$
    relative to the incident edges of $r$.}
    \label{pf121}
\end{figure}


    (2) Let $(r,i)$ be an available type for $c$ and $K\in \cK_{<\alpha}$ is admissible for $v_1$ in $H_1=H_{r,i}(v_1)$. So, $\vis(v_1)\cap K=\emptyset$, and hence $v_1\notin R_K$ using \Cref{lem:avoid}. If $v_2\in R_K$, then there must be an edge $e$ of $R_K$ such that $v_1$ and $v_2$ lie on the either side of the supporting line of $e$. As $e\in L_3^{(\alpha)}$, $v_1$ and $v_2$ lie in different cells of $C_\alpha$, which is a contradiction. So, $v_2 \notin R_K$, and hence $\vis(v_2)\cap K=\emptyset$ using \Cref{lem:avoid}. Now, we have $K\subset H_1$. If $H_1\subseteq H_2=H_{r,i}(v_2)$, then we are done, since $K\subset H_1\subseteq H_2$. Otherwise, let $H_2\subsetneq H_1$. Let $K=H_{r',i'}(v)$ for some point $v$, a reflex vertex $r'$ and $i'\in \{1,2\}$. Using \Cref{lem:contain}, $r'\in A(H_1)$. Observe that, $A(H_1)\setminus A(H_2)$ is the polygonal path ($\subset \bd(\pol)$) between $h_r(v_1)$ and $h_r(v_2)$ which does not contain $r$. Suppose, for contradiction, we assume that, $r'\notin A(H_2)$. Then, $v_1,v_2$ lie on the either side of the line $l'$ through $r$ and $r'$ (see \Cref{pf122}). 

    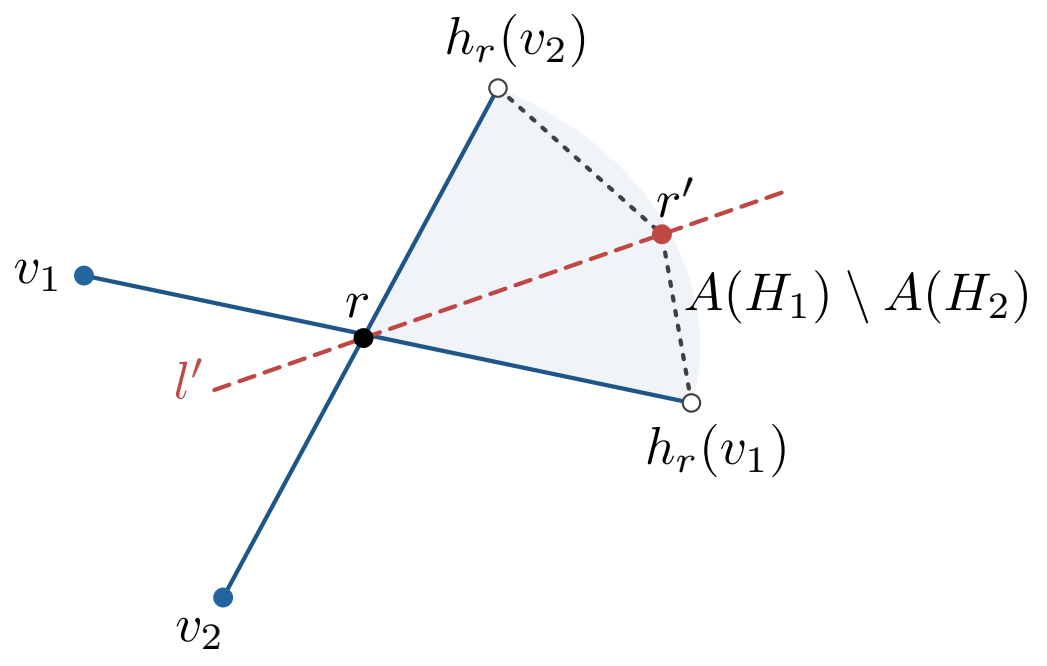
\begin{figure}[ht!]
    \centering
\begin{tikzpicture}[scale=0.78,line cap=round,line join=round,
    every node/.style={font=\small}]
  \definecolor{proofblue}{RGB}{34,101,159}
  \definecolor{proofred}{RGB}{192,72,68}
  \coordinate (r) at (0,0);
  \coordinate (hTwo) at (1.14,2.12);
  \coordinate (hOne) at (2.78,-0.55);
  \coordinate (rp) at (2.53,0.88);
  \coordinate (vone) at (-2.37,0.53);
  \coordinate (vtwo) at (-1.19,-2.20);
  \fill[proofblue!7] (r) -- (hTwo) .. controls (2.70,1.49)
      and (3.03,0.16) .. (hOne) -- cycle;
  \draw[proofblue!85!black,thick] (vtwo) -- (hTwo);
  \draw[proofblue!85!black,thick] (vone) -- (hOne);
  \draw[black!75,dotted,thick] (hTwo) -- (rp) -- (hOne);
  \draw[proofred,densely dashed,thick]
     ($(r)!-0.50!(rp)$) -- ($(r)!1.43!(rp)$);
  \node[proofred, left=4pt] at ($(r)!-0.39!(rp)$) {$l'$};
  \fill (r) circle (2.4pt);
  \fill[proofred] (rp) circle (2.4pt);
  \fill[proofblue] (vone) circle (2.4pt);
  \fill[proofblue] (vtwo) circle (2.4pt);
  \draw[black!75,fill=white] (hTwo) circle (2.1pt);
  \draw[black!75,fill=white] (hOne) circle (2.1pt);
  \node[above =0.1pt] at (-0.05,0) {$r$};
  \node[above =0.1pt] at (2.65,0.85) {$r'$};
  \node[left=0.1pt] at (vone) {$v_1$};
  \node[below left=0.1pt] at (-1,-2.20) {$v_2$};
  \node[above=0.1pt] at (1.3,2.12) {$h_r(v_2)$};
  \node[below =0.1pt] at (3,-0.55) {$h_r(v_1)$};
  \node[align=left,right=2pt] at (2.45,0.35)
    {$A(H_1)\setminus A(H_2)$};
\end{tikzpicture}
    \caption{A reflex vertex $r'$ on the boundary
    arc $A(H_1)\setminus A(H_2)$.}
    \label{pf122}
\end{figure}


    So, $v_1$ and $v_2$ lie in different cells of $C_\alpha$, since $l'\in L_2$. This is a contradiction. Therefore, $r'\in A(H_2)$, and hence $K\subseteq H_2$ using \Cref{lem:contain}. Thus, $K\in \cK_{<\alpha}$ is admissible for $v_2$ in $H_{r,i}(v_2)$. The reverse direction is symmetric to the forward direction.
\end{proof}


\paragraph*{Obtaining a suitable family:} 
Let $c\in C_\alpha$ and $(r,i)$ be an available type for $c$. By \Cref{lem:const}, set of all admissible entries are same for all points of $iN(c)$. Thus, we define, \emph{$\cQ_{\alpha,c}$} $\coloneq \{K\in \cK_{<\alpha}~|~K\text{ is admissible for $x$ in}\text{ $H_{r,i}(x)$ for some $x\in iN(c)$}\}$. For each $K\in \cQ_{\alpha,c}$ and $x\in iN(c)$, $A(K)\subsetneq A(H_{r,i}(x))$ 
since $K\subsetneq H_{r,i}(x)$, and $A(K)$ is a simple polygonal path. So, each $A(K)$ can be represented as an interval on $x$-axis by a suitable parametrization. Also, by \Cref{lem:arc}, admissible entries are disjoint exactly when their arcs are. Thus, $\cQ_{\alpha,c}$ can be represented as a set $I_{\alpha,c}$ of intervals on $x$-axis, where let $l_K$ be the interval corresponding to the entry $K\in \cQ_{\alpha,c}$. Assign weight $\lambda_K=\nw(s)$ to $l_K$, where $K$ is the entry of the state $s$. Thus we get a family \emph{$\cS_{\alpha,c}$} $\subseteq \cQ_{\alpha,c}$ by solving weighted interval scheduling problem with input $I_{\alpha,c}$ and weight $\lambda_K$ for $l_K\in I_{\alpha,c}$. Running time for obtaining $\cS_{\alpha,c}$ is $\oo(|\cQ_{\alpha,c}|\log (|\cQ_{\alpha,c}|))=\oo(\rho n\log(\rho n))=\oo(\rho n\log n)$. The following observation follows directly from \Cref{thm:rec} and the above argument. 

\begin{observation}\label{obs:equivcheck}
    For $x\in c$, $H_{r,i}(x)$ works for $s=(\alpha,r,i)$ if and only if $1+\sum_{K\in \cS_{\alpha,c}}\lambda_K\geq \alpha$.
\end{observation}




\begin{lemma}\label{lem:cell}
    Let $c\in C_\alpha$ and $(r,i)$ be an available type for $c$. Then, there exists a vertex $v_{_{r,i}}^{c}~(\neq r)$ of the cell $c$ such that $H_{r,i}(v_{_{r,i}}^{c})\subseteq H_{r,i}(x)$ for each $x\in c$.
\end{lemma}
\begin{proof}
    For any $y\in c$, let $l_y$ be the line through $r$ and $y$. The cap of any point on $l_y\cap c$ associated with $(r,i)$ is equal to $H_{r,i}(y)$. Now $c\setminus l_y$ consists of two connected convex regions $R_1^y$, and $R_2^y$. Let $R_1^y$ be the region such that all the points of $R_1^y$ and the edges incident to $r$ lie on the same side of $l_y$. Then for any point $y_i\in R_i^y$, we have $H_{r,i}(y_1)\subseteq H_{r,i}(y)\subseteq H_{r,i}(y_2)$. Observe that, there exist a vertex $v$ of the cell $c$ such that $R_1^y=\emptyset$, since there are exactly two tangents from $r$ to $c$ and if we choose a vertex responsible for the suitable tangent (the tangent such that the cell and the edges incident to $r$ lie on the either side of that tangent) as $y$, then $R_1^y$ is empty and $R_2^y$ consists of the whole $c\setminus l_y$ Call that vertex $v_{r,i}^c$. So, we have $H_{r,i}(v_{r,i}^c)\subseteq H_{r,i}(x)$ for each $x\in c$.
\end{proof}

\begin{algorithm}[ht!]
\caption{Cap table and root records}\label{alg:main}
\KwIn{a simple polygon $\pol$ whose vertices are in general position}
\KwOut{$\mws(\pol)$, the state records, and the root-cell records}
\lIf{$\rho=0$}{\Return $1$, with the average of the vertices as the witness}
Set every state to $\emptyset$\;
\For{$\alpha=1,\dots,h$}{
  Build $\cA_\alpha$ from the entries of rows below $\alpha$\;
  \ForEach{cell $c\in C_\alpha$}{
    \ForEach{type $(r,i)$ available at $g(c)$}{
      Find the admissible lower entries $\cQ_{\alpha,c}$ and a best disjoint family $\cS_{\alpha,c}$\;
      \If{$1+\sum_{K\in \cS_{\alpha,c}}\lambda_K\ge \alpha$}{
        Find the extreme corner $v_{_{r,i}}^{c}$ and $H_{r,i}(v_{_{r,i}}^{c})$\;
        \lIf{$H_{r,i}(v_{_{r,i}}^{c})$ is non-empty and smaller than the current entry of $(\alpha,r,i)$}{store it with $c$, $v_{_{r,i}}^{c}$ and a child record of weight $\alpha-1$}
      }
    }
  }
  Compute $R_K$ for the new entries\;
}
Build $\cA_{\mathrm{root}}$ by the same rules from all entries\;
\ForEach{open cell $c$ of $\cA_{\mathrm{root}}$ inside $\pol$}{
  Cut $\bd(\pol)$ at a vertex visible from $g(c)$\;
  Schedule the entries avoided by $g(c)$; set $b(c)=1+{}$weight\;
}
\Return $\max_cb(c)$ and the records\;
\end{algorithm}

\paragraph*{Brief Description of \Cref{alg:main}.}
The algorithm first handles the special case when the polygon is convex, i.e., has no reflex vertex (line 1) and initializes all states as empty (line 2). It then processes the states in increasing order of $\alpha$, constructing an arrangement from the entries of previously computed rows (lines 3–4) using $R_K$ computed in line 11. For each cell and each available type, it finds a maximum-weight collection $\cS_{\alpha,c}$ of pairwise disjoint admissible entries using weighted interval scheduling (lines 5–7). If these entries can form a witness set of the required size, it updates the corresponding state with the smallest cap and stores the information needed to reconstruct the witness set explicitly (lines 8–10). It then computes the regions associated with the newly obtained entries (line 11). Finally, it constructs an arrangement for the root, evaluates each cell to find the largest possible witness set, and returns its size along with the corresponding records (lines 12–16).

\begin{lemma}\label{thm:table}
After completing row $\alpha$ of \Cref{alg:main}, the entry for $s=(\alpha,r,i)$ is exactly $K^*(s)$, and its
child record is admissible throughout its recorded cell.
\end{lemma}

\begin{proof}
By induction on $\alpha$. By \Cref{thm:rec} and \Cref{lem:const}, a cell contributes working caps to a state
exactly when its schedule is heavy enough, and \Cref{lem:cell} computes the infimum over that
cell. Caps of one type are nested, so the smallest cell value is the intersection over all contributing cells.
\end{proof}
\begin{lemma}\label{thm:value}
\Cref{alg:main} returns $\mws(\pol)$.
\end{lemma}

\begin{proof}
For every root cell, the chosen entries (after interval scheduling) are disjoint and avoided by $g(c)$. So, using \Cref{obs:realize}, we get $b(c)$ witnesses, and $b(c)\le\mws(\pol)$. Thus, $\max_cb(c)\le \mws(\pol)$. Conversely, hang the tree of a largest witness set from a witness $w$. The caps two levels below $w$ work for defined states, have total weight $\mws(P)-1$, are disjoint, and miss $\Vis(w)$. Their entries do as well. The sets $R_K$ are closed and miss $w$, so $w$ can be moved into an open cell of $\cA_{\mathrm{root}}$ that keeps every avoidance. By \Cref{lem:const} the representative of that cell admits the same family.
\end{proof}

\begin{proof}[Proof of \Cref{thm:main}.]
    If $\rho=0$, the polygon is convex, every point sees all of $\mathcal{P}$, and one witness is the best possible, which is what line~1 returns. Now let $\rho\geq 1$. By \Cref{lem:size}, no witness set has more than $h=\lfloor n/3\rfloor$ points, so the rows $\alpha=1,\dots,h$ cover every possible size. The rows are filled in increasing order of~$\alpha$. By \Cref{thm:table}, after row~$\alpha$ is finished, the entry of every state $s=(\alpha,r,i)$ is exactly the smallest cap $K^*(s)$. Together with this entry, line~10 stores a cell, an extreme corner (\Cref{lem:cell}), and a child record (\Cref{l1}): a family of lower states whose entries are pairwise disjoint, are admissible throughout the stored cell (\Cref{lem:const}), and have total weight exactly $\alpha-1$. By \Cref{thm:value}, the value $\max_c b(c)$ returned in line~16 equals $\mu_{\mathrm{WS}}(\mathcal{P})$. To output actual points, we start from a root cell~$c$ with the largest value $b(c)$ and place a witness at $g(c)$. The entries chosen for this cell are pairwise disjoint and are not seen by $g(c)$ (\Cref{lem:avoid}). For each chosen state, we place one witness inside its stored cell, very close to its stored corner. The cap of this point is then very close to the entry $K^*(s)$, so it stays disjoint from its siblings, stays unseen by its parent, and stays inside its parent's cap; the reconstruction section shows that such points exist. We then do the same for the child record of this state, and so on down to row~$1$. At every level, \Cref{obs:realize} shows that the points chosen so far form a witness set. A state of weight~$\alpha$ contributes its own witness and, through its child record, $\alpha-1$ more, so the final set has exactly $b(c)=\mu_{\mathrm{WS}}(\mathcal{P})$ points.

    (Running Time) The dynamic programming table has $\oo(h)=\oo(\lfloor n/3\rfloor)$ rows. Before row~$\alpha$ is filled, the table holds at most $2\rho h=\oo(\rho n)$ entries, and each set $R_K$ has $\oo(n)$ edges. So the arrangement $\mathcal{A}_\alpha$ has $\oo(\rho n^2)$ lines, it can be built in $\oo(\rho^2 n^4)$ time, and it has $\oo(\rho^2 n^4)$ cells. In each cell we do three things. First, we compute the visibility region of $g(c)$ in $\oo(n)$ time. This region tells us which types are available, and there are at most $\rho$ of them. Second, we check which earlier entries $K$ are avoided by $g(c)$. Each $R_K$ is prepared for point location when it is created in line~11, so one check takes $\oo(\log n)$ time, and all checks in a cell take $\oo(\rho n\log n)$ time. These checks do not depend on the type, so we do them only once per cell. Third, for each available type, we test containment with \Cref{lem:contain} in constant time per entry, and then solve one weighted interval scheduling problem with $\oo(\rho n)$ intervals. Sorting the intervals makes this cost $\oo(\rho n\log n)$. The third step is the most expensive one: it costs $\oo(\rho^2 n\log n)$ per cell, and therefore $\oo(\rho^4 n^5\log n)$ per row. Lines~8--11 cost less than this. Over all $h=\oo(n)$ rows, the total is $\oo(\rho^4 n^6\log n)$. The final phase in lines~12--16 visits $\oo(\rho^2 n^4)$ cells and solves one scheduling problem in each. It costs $\oo(\rho^3 n^5\log n)$, so it does not change the bound.
\end{proof}

\paragraph*{AI Disclosure.}
ChatGPT was used to assist with language editing, polishing, manuscript organization, figure preparation, and improving the clarity of exposition. Additionally, ChatGPT provided limited technical assistance in the development  of the dynamic programming approach. All geometric characterizations were developed independently by the authors, with most of them established in our earlier work \cite{DBLP:journals/corr/abs-2605-01592}.  The authors take full responsibility for the correctness, originality, and scientific content of the manuscript.

\paragraph*{Acknowledgment:} The authors would like to thank Bodhayan Roy for several helpful discussions.

\bibliographystyle{alpha}
\bibliography{cite}

\end{document}